# Sampling-based Buffer Insertion for Post-Silicon Yield Improvement under Process Variability

Grace Li Zhang, Bing Li and Ulf Schlichtmann
Institute for Electronic Design Automation, Technische Universität München, Munich, Germany
Email: {grace-li.zhang, b.li, ulf.schlichtmann}@tum.de

*Abstract*—At submicron manufacturing technology nodes process variations affect circuit performance significantly. This trend leads to a large timing margin and thus overdesign to maintain yield. To combat this pessimism, post-silicon clock tuning buffers can be inserted into circuits to balance timing budgets of critical paths with their neighbors. After manufacturing, these clock buffers can be configured for each chip individually so that chips with timing failures may be rescued to improve yield. In this paper, we propose a sampling-based method to determine the proper locations of these buffers. The goal of this buffer insertion is to reduce the number of buffers and their ranges, while still maintaining a good yield improvement. Experimental results demonstrate that our algorithm can achieve a significant yield improvement (up to 35%) with only a small number of buffers.

## I. INTRODUCTION

At advanced technology nodes, process variations have become relatively larger, and thus caused expensive overdesign to guarantee yield. To alleviate the effect of process variations, many researchers have introduced special circuit components and mechanisms. For instance, post-silicon tuning components can be inserted into the circuit to alleviate the effect of process variations. Since physical parameters are fixed for each individual chip after manufacturing, tuning at the post-silicon stage can improve the performance specifically for each chip.

A widely used post-silicon tuning technique is clock tuning using delay buffers. For example, the structure of the delay buffer used in [1] is illustrated in Fig. 1. The delay of such a buffer can be changed by setting the configuration bits in the three registers. In high-performance designs, these tuning buffers are inserted during the design phase. After manufacturing, the delay values of these buffers are configured to allot critical paths more timing budget by shifting clock edges toward the stages with smaller combinational delays. These critical paths might be different in individual chips due to process variations, so that only post-silicon tuning can counterbalance them effectively. With this post-silicon tuning, chips that might have failed to meet timing specifications can be revitalized, leading to an increased yield at the expense of additional area required by these buffers.

In this paper, we propose a method to insert post-silicon tuning buffers at the design phase for yield improvement. This method uses Monte Carlo simulation to emulate produced chips and confine tuning buffers to as few flip-flops as possible in each sample. After the whole simulation is finished, only those buffers that are critical to the yield are kept in the circuit. The proposed method captures the locations and ranges of tuning buffers directly without an intermediate formulation so that heuristics in statistical optimization are avoided. In addition, this direct simulation method provides correlation information between buffer tuning values. This information is used to group buffers so that the total area taken by the buffers is reduced effectively.

The rest of this paper is organized as follows. In Section II we give an overview of timing constraints for circuits with post-silicon clock tuning buffers and formulate the buffer insertion problem. We explain the proposed method in detail in Section III. Experimental results are shown in Section IV. Conclusion and future work are given in Section V.

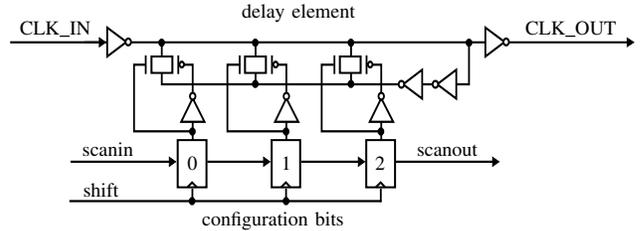

Fig. 1. Post-silicon tuning buffer in [1].

## II. TIMING CONSTRAINTS WITH CLOCK BUFFERS AND PROBLEM FORMULATION

The timing constraints with clock tuning buffers can be explained using Fig. 2, where two flip-flops with such buffers are connected by a combinational circuit. Assume that the clock signal switches at reference time 0. Then the clock events at flip-flops $i$ and $j$ happen at time $x_i$ and $x_j$, respectively. To meet the setup time and hold time constraints, the following inequations must be satisfied

$$x_i + \overline{d}_{ij} \leq x_j + T - s_j \tag{1}$$
$$x_i + \underline{d}_{ij} \geq x_j + h_j \tag{2}$$

where $x_i$ and $x_j$ are delay values of tuning buffers, $\overline{d}_{ij}$ ($\underline{d}_{ij}$) is the maximum (minimum) delay of the combinational circuit between flip-flops $i$ and $j$, $s_j$ ($h_j$) is the setup (hold) time of flip-flop $j$, and $T$ is the clock period. Here the clock buffers introduce two delay variables into the constraints (1) and (2). Without them, the two inequations fall back to the classic timing constraints of digital circuits.

Owing to area constraints, the configurable delay of a clock buffer usually has a limited range. This range determines the size of the buffer so that it should be as small as possible. Assume that the lower bound of the tuning values of buffer $i$ is $r_i$ and the range of the buffer is $\tau_i$. The delay value of buffer $i$ can thus be constrained by a range window as

$$r_i \leq x_i \leq r_i + \tau_i. \tag{3}$$

Unlike [2], here we model the range window of the tuning values as asymmetrical with respect to 0 to achieve a maximal flexibility. Furthermore, $x_i$ may only take discrete values due to implementation issues.

When process variations are considered, the combinational delays $\overline{d}_{ij}$ and $\underline{d}_{ij}$, setup time $s_j$ and hold time $h_j$ in (1) and (2) become random variables. In addition, the tuning delays $x_i$ and $x_j$ also become statistical because the clock buffers are subject to process variations. Since these variations can be decomposed and merged with $\overline{d}_{ij}$, $\underline{d}_{ij}$, $s_j$ and $h_j$, e.g., using the canonical form in [3], we assume henceforth that a tuning delay can be configured to a fixed value in (3) for simplicity.

In real circuits, the number of clock tuning buffers is limited. To maintain a good yield improvement, the problem of buffer insertion can thus be described as

**Problem Buffer-Insertion:**
*Select a set of flip-flops to insert clock tuning buffers to improve the yield of the circuit. The number and ranges of buffers should be kept as small as possible. In addition, the bounds of tuning ranges should be determined, which can vary at different buffers and cover negative delays.*

The predominant challenge in solving the Buffer-Insertion problem comes from the random variables in (1) and (2). These variables make it very difficult to model the formulation above as an optimization problem, since statistical optimization very often resorts to heuristic approximations based on Monte Carlo simulation such as in [2]. In addition, the variables $x_i$ and $x_j$ in (1) and (2) may only take discrete values in the range window defined by (3). For example, a de-skew buffer in [4] can be configured to only 20 discrete delays. In this case, integer linear programming (ILP) becomes almost the only method available to deal with the constraint set defined by (1)–(3). In the proposed method, we use ILP-based Monte Carlo simulation to identify the buffer locations and tuning ranges directly.

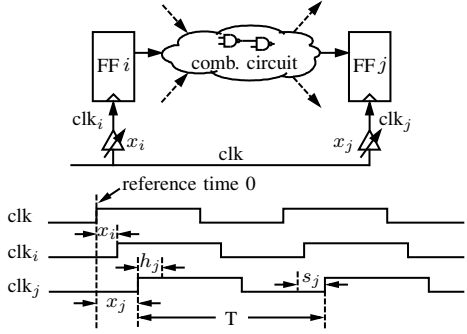

Fig. 2. Timing of circuits with tuning buffers.

### III. SOLVING THE BUFFER-INSERTION PROBLEM BY SAMPLING-BASED SIMULATION

To identify where buffers should be inserted, we use a sampling-based method to deal with the complexity from process variations. The overall flow of the proposed method is shown in Fig. 3. The input of the proposed method includes the circuit structure, statistical gate delays and specifications of available clock buffers such as the given maximum range and the number of discrete steps. The proposed method improves the yield of the circuit with respect to a give clock period $T$, the same as in [2]. The output of the proposed method are locations of buffers and the reduced ranges of their tuning delays. The number and ranges of buffers should be kept as small as possible.

In the proposed method, we first assume that each flip-flop has a tuning buffer attached. We then sample the statistical variables in (1) and (2) and minimize the number of buffers required to achieve the given clock period for a specific sample. Here a sample can be considered as a representative chip after manufacturing. With enough samples, the trend where buffers should be inserted can thus be recognized, and only those buffers that affect the yield significantly are kept in the circuit.

The proposed method contains three major steps. In the first step, we allow the lower bounds of buffers to float freely, because at this stage the lower bound $r_i$ in (3) is still unknown. For each sample, we minimize the number of required buffers and push the tuning values of buffers toward 0. Because these tuning values should be covered by the given tuning range windows, this value concentration can narrow the distributions of tuning values effectively. At the end of this step, the lower bound $r_i$ is determined by finding the range window with a range $\tau_i$ that covers the most tuning values in the simulation to maintain the yield.

With the lower bounds determined, we rerun the sampling process in the second step. For each sample, we minimize the number of required buffers and then concentrate tuning values toward their average. The latter is different from the previous step. In the first step, the lower bounds of buffers float freely, so that an average of tuning values does not reflect the center of a tuning range. Therefore, we focus on the reduction of buffer number by pushing tuning values toward 0. In the second step, the bounds of buffer ranges have been determined, so that the average can represent the trend of tuning values. Consequently, pushing tuning values toward this average may reduce the ranges of buffers. Finally, the ranges of buffers are determined by the smallest and the largest tuning values,

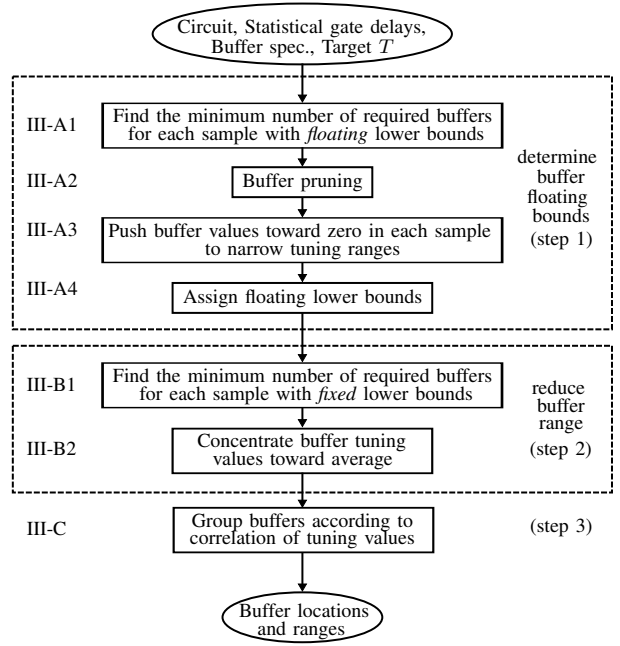

Fig. 3. Overall flow to determine buffer locations and tuning ranges. The numbers are the sections in which the corresponding steps are described.

respectively.

In the last step, buffers are grouped according to their tuning correlation. If two buffers have highly correlated tuning values, they can share the same buffer to save area. The correlation information is a natural result of the sampling-based method so that buffers can be grouped easily.

The advantage of the simulation-based method is that it does not rely on the distributions of random variables. The challenge of this method is a large runtime, although it can be parallelized easily onto multiple CPU cores for acceleration. To solve the runtime problem, we introduce several techniques which can accelerate the proposed method effectively.

### A. Determining lower bounds of the range windows

In this step, we have no information about where to insert buffers and how to set the lower bounds of these buffers. Since the circuit structure and delay distributions all affect the locations of buffers, we resort to sampling-based simulation to gather preliminary information. The basic idea is that we sample the delays of the circuit and create an ILP model for each sample to confine the tunings of buffers to as few flip-flops as possible, while the lower bounds of the tuning ranges are allowed to float freely. With a large number of samples, the trend where the buffers should be inserted can be exposed by the concentration of tunings to only a few flip-flops. These flip-flops are actually the critical ones affecting the yield of the circuit.

To model whether a buffer should be adjusted, we use a binary variable $c_i$ assigned for the $i$th buffer and defined as

$$c_i = \begin{cases} 1 & \text{the } i\text{th buffer is adjusted, so that } x_i \neq 0, \\ 0 & \text{otherwise when } x_i = 0. \end{cases} \quad (4)$$

According to [5], this constraint can be transformed to

$$x_i \leq c_i \Gamma \quad (5)$$
$$-x_i \leq c_i \Gamma \quad (6)$$

where $\Gamma$ is very large constant. If $x_i$ is larger than 0, then $c_i$ must be 1. If $x_i$ is smaller than 0, $c_i$ should also be set to 1. The only situation that $c_i$ can be set to 0 is when $x_i$ is 0, meaning that there is no tuning.

From the analysis above, we can observe that $c_i$ is an upper bound of the tuning number for the $i$th buffer, since $c_i$ can also be set to 1 even if $x_i = 0$. Accordingly, the sum of $c_i$s for all tuning buffers is an upper bound of the number of required buffers for the sample. This upper bound can be expressed as

$$c_{sum} = \sum_i c_i, \, i \in I \quad (7)$$

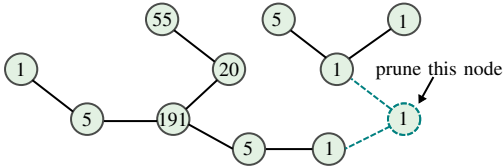

Fig. 4. Pruning buffers with low tuning numbers.

where $I$ is the index set of all the tuning buffers. If we minimize this upper bound, the total number of tunings is also minimized.

Besides minimizing the total number of buffers $c_{sum}$, we also hope that the tuning values are not scattered widely, so that they can be covered by a range window with the range equal to $\tau_i$. To achieve this goal, we minimize the distance between these values and 0. Consequently, the second objective of the optimization task is $\sum_{i=1}^{I} |x_i|$, which is total distance of the tuning values to 0.

In our formulation, the number of buffers has a priority in optimization, because fewer buffers mean a smaller area as well a simpler layout design. Therefore, we split the optimization problem into two. In the first one we set the objective to $c_{sum}$ to find the minimum number of required buffers for each sample. This number is used as a constraint in the second optimization problem to minimize $\sum_{i=1}^{I} |x_i|$. After the first optimization, we prune buffers with few tunings so that the execution time of solving the second optimization problem can be reduced significantly.

*1) Minimizing buffer tuning with floating lower bounds*

With the number of tunings $c_{sum}$ defined in (7), we describe the first optimization problem as follows.

$$\text{For each sample } m_k \in \mathcal{M}, \tag{8}$$
$$\text{Minimize} \quad c_{sum} \tag{9}$$
$$\text{s.t. setup and hold time constraints (1) and (2)} \tag{10}$$
$$\quad \text{constraints of required tuning buffers (5)–(7)} \tag{11}$$
$$\quad \text{range constraints (3)} \tag{12}$$
$$\quad r_i \leq 0 \text{ and } r_i + \tau_i \geq 0 \tag{13}$$

where $m_k$ represents the $k$th sample of the circuit. $\mathcal{M}$ is the set of all samples in the simulation and its cardinality $|\mathcal{M}|$ is the number of samples in the simulation. In the formulation above, the lower bounds of $r_i$ are variables and determined by the solver. The constraint (13) requires that 0 is covered by the range window. After the optimization problem (8)–(13) is solved for the $k$th sample, we denote the tuning number of the whole circuit in the sample $m_k$ by $n_k$.

*2) Buffer pruning*

Since the optimization problem above tries to reduce the number of required buffers as much as possible, many buffers are not adjusted or adjusted only for a few times even with the floating lower bounds. Fig. 4 shows such an example, where nodes represent buffers and edges represent combinational connections between flip-flops. The numbers represent how many times in the $|\mathcal{M}|$ simulation samples the corresponding buffer is used. We remove the nodes whose tuning numbers are no large than one and are not connected to other critical nodes. The latter is defined as the nodes with tuning numbers no smaller than a given number, which is set to 5 as the number of simulation samples is set to 10 000 in our experiments. For example, we remove the node with dashed line from the graph in Fig. 4. This removal not only accelerates the following steps due to fewer nodes in the graph, but may also reduce the problem space significantly by partitioning the graph into unconnected sub-graphs.

*3) Concentrating tuning values around zero*

The optimization problem in (9)–(13) only reduces the number of tunings in each sample. Since the optimization problems for different samples are solved separately, the solver does not guarantee to use the same set of buffers in case more than one feasible solution exists. Consequently, the tuning values of a buffer from all the samples may become scattered into a wide distribution, as illustrated in Fig. 5a, where the x-axis represents the tuning delays of a buffer in all simulation samples, and the y-axis the number of occurrences of discrete delay values.

To push the scattered tuning values into a narrower range, we minimize their absolute values in the optimization. In this way, the solver tries to return the buffer values around 0 as much as possible, while maintaining the minimum tuning number $n_k$ of buffers in the $k$th simulation sample $m_k$ obtained by solving the problem (8)–(13). The latter constraint guarantees that the number of tuning is still optimal for each sample.

$$\text{For each sample } m_k \in \mathcal{M}, \tag{14}$$
$$\text{Minimize} \quad \sum_{i=1}^{I} |x_i| \tag{15}$$
$$\text{s.t. constraints (10)–(13)} \tag{16}$$
$$\quad c_{sum} \leq n_k \tag{17}$$

where the objective function (15) can be transformed into a linear form easily [5].

*4) Assigning floating lower bounds*

After tuning values are pushed toward 0, we try to cover all the tuning values using a range window of the maximum range $\tau_i$. As shown in Fig. 5b, the range window slides along the x-axis. Since the y-axis represents the numbers of the corresponding tuning values occurred in all simulation samples, the overall covered buffer tunings by the window is thus the sum of the tuning numbers in the window. For yield improvement, we select the range window that covers the largest number of tuning values. In this way, the lower bound of the tuning range is determined as the leftmost value of the range window.

### B. Identify buffers with fixed lower bounds

With the lower bounds of buffers determined, the locations and ranges of buffers should be evaluated again since the fixed lower bounds for all the buffers invalidate previous simulation results. The basic concept of learning from simulation is similar to the steps above.

*1) Minimizing buffer tunings with fixed range bounds*

After the lower bounds are fixed, the variables $r_i$ in (3) and (13) become constants. Therefore, we execute the simulation-optimization process (8)–(13) again to capture more realistic tuning values of buffers. In practice, this simulation step can be skipped if the number of tunings outside the determined range windows is very small. In our method, we skip this simulation step if the missing tunings appear in less than 0.1% of all simulation samples.

*2) Concentrating buffer tunings toward the average*

After running simulation samples with fixed buffer ranges, the tuning values are all confined in given range windows. Their values, however, may be scattered widely because the solver only returns feasible delays without considering the results of previous tuning values. This is the same problem described in Section III-A3. Similarly, we try to centralize the buffer values further toward the average tuning delays $x_{avg,i}$ of all the samples. The mathematical problem is formulated below. Here we also use the minimum tuning numbers $n_k$ to guarantee that in this new round of simulation the solver does not increase the number of buffers.

$$\text{For each sample } m_k \in \mathcal{M}, \tag{18}$$
$$\text{Minimize} \quad \sum_{i=1}^{I} |x_i - x_{avg,i}| \tag{19}$$
$$\text{s.t. constraints (10)–(13)} \tag{20}$$
$$\quad c_{sum} \leq n_k \tag{21}$$

This step is very similar to the problem formulation in (14)–(17). The only difference is that we centralize the delay tunings toward the average instead of zero. In (14)–(17) we try to find the most effective lower bounds of buffers so that we use as few buffers as possible. Consequently, we concentrate the tuning values around zero. In (18)–(21) the lower bounds have been fixed and we only try to narrow the scattered tuning values to reduce the ranges of the buffers. Therefore, we push all tuning values toward the average instead of zero. Finally, the ranges of buffers are assigned according to the largest and the smallest tuning value, as illustrated in Fig. 5c.

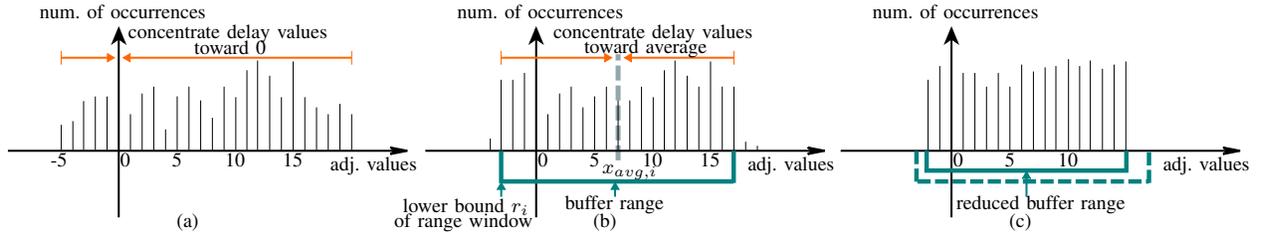

Fig. 5. Concentrating tunings of a buffer in all samples.

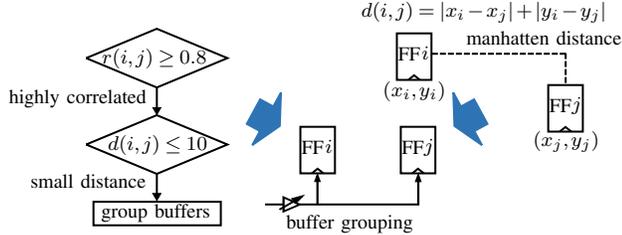

Fig. 6. Buffer grouping according to tuning correlation and distance. Correlation threshold $r_t$ is set to 0.8. Distance threshold $d_t$ between buffers is set to ten times of the minimum distance between flip-flops.

### C. Grouping

In the last step of buffer insertion, we group buffers with similar tuning values to reduce the number of buffers inserted into the circuit. Buffers in the same group are implemented by only one physical buffer and the tuning values are shared by all the flip-flops connected to the buffer. The concept of grouping is illustrated in Fig. 6.

In grouping buffers, we first calculate the correlation coefficients of tuning values of individual buffer pairs. If the mutual correlation coefficients between several buffers are all above the threshold $r_t$ and their distance is smaller than $d_t$, they are grouped together and implemented with only one physical buffer. In practice, designers can also constrain the total number of buffers in the circuit. If the buffer number after grouping still exceeds the specified number, we remove the buffers with the fewest tunings until the number of buffers meets the specification.

### IV. EXPERIMENTAL RESULTS

The proposed method was implemented in C++ and tested using a 3.20 GHz CPU with one thread. We demonstrate the results with circuits from the ISCAS89 benchmark set and from the TAU 2013 variation-aware timing analysis contest. To these circuits we also added clock skews so that they have more critical paths. Information about these circuits is shown in Table I, where $n_s$ is the number of flip-flops and $n_g$ the number of logic gates. We assumed that the maximum allowed buffer ranges were 1/8 of the original clock period [4]. All tuning delays are assumed discrete with 20 steps. The logic gates in the circuits were mapped to a library from an industry partner. The standard deviations of transistor length, oxide thickness and threshold voltage were set to 15.7%, 5.3% and 4.4% of the nominal values. The ILP solver for the optimization problems was Gurobi [6]. In the proposed method, we generate 10 000 samples to capture the locations and ranges of buffers.

The experiment results are shown in Table I. To test a circuit, we first run Monte Carlo simulation to calculate the mean $\mu_T$ and standard $\sigma_T$ of the clock period without post-silicon tuning buffers. The original yields $Y_o$ of the circuits with the clock period equal to $\mu_T$, $\mu_T + \sigma_T$ and $\mu_T + 2\sigma_T$ are thus about 50%, 84.13%, 97.72%, respectively. In Table I the columns $Y(\%)$ shows the yields with post-silicon tuning buffers. The columns $Y_i(\%)$ show the yield improvements compared with the original yields without buffers, equal to $Y - Y_o$. From this comparison, we can see clearly that the yields of circuits can be improved significantly (up to 35.97%).

The numbers of buffers inserted into the circuits to achieve yield improvements above are shown in the columns $N_b$. These numbers are less than 1% of the numbers of flip-flops in the circuits for a significant yield improvement. In addition to the numbers of buffers, the average ranges of buffers are shown in the columns $A_b$. Since we centralize the buffer values as illustrated in Fig. 5c, the ranges of buffers are much smaller than the maximum buffer range 20 so that the area taken by inserted buffers can be reduced.

In Table I the columns $T(s)$ show the runtimes of the proposed method in different settings. For the largest circuit pci_bridge32, the proposed method needs 5124.25 seconds to finish the computation. These runtimes are acceptable since buffer insertion is normally executed at a late design phase only for a few times.

### V. CONCLUSION

In this paper, we propose a sampling-based method to determine locations and ranges of post-silicon tuning buffers in a circuit to improve yield. Experimental results confirm that yield can be improved significantly with a small number of buffers. Future work includes post-silicon testing and configuration of delays buffers to achieve the given clock period. Challenges are a balance between testing cost and yield improvement in complex scenarios such as clock binning.

TABLE I. RESULTS OF BUFFER NUMBER AND YIELD IMPROVEMENT

| Circuit | | | $\mu_T$ | | | | | $\mu_T + \sigma_T$ | | | | | $\mu + 2\sigma_T$ | | | | |
|---|---|---|---|---|---|---|---|---|---|---|---|---|---|---|---|---|---|
| | $n_s$ | $n_g$ | $N_b$ | $A_b$ | $Y(\%)$ | $Y_i(\%)$ | $T(s)$ | $N_b$ | $A_b$ | $Y(\%)$ | $Y_i(\%)$ | $T(s)$ | $N_b$ | $A_b$ | $Y(\%)$ | $Y_i(\%)$ | $T(s)$ |
| s9234 | 211 | 5597 | 2 | 12.50 | 77.11 | 27.11 | 54.22 | 2 | 12.00 | 95.94 | 11.81 | 47.11 | 2 | 11.00 | 99.18 | 1.46 | 7.79 |
| s13207 | 638 | 7951 | 5 | 9.80 | 72.37 | 22.37 | 156.05 | 5 | 14.20 | 96.42 | 12.29 | 92.84 | 6 | 17.30 | 99.53 | 1.81 | 24.16 |
| s15850 | 534 | 9772 | 5 | 19.80 | 69.34 | 19.34 | 223.09 | 5 | 19.40 | 94.33 | 10.20 | 90.89 | 5 | 15.20 | 99.12 | 1.40 | 23.42 |
| s38584 | 1426 | 19253 | 11 | 9.74 | 85.97 | 35.97 | 1800.14 | 7 | 13.14 | 98.48 | 14.35 | 683.62 | 7 | 13.57 | 98.94 | 1.22 | 223.95 |
| mem_ctrl | 1065 | 10327 | 10 | 11.90 | 67.11 | 17.11 | 1206.54 | 10 | 11.70 | 94.58 | 10.45 | 531.78 | 10 | 8.70 | 98.91 | 1.19 | 147.89 |
| usb_funct | 1746 | 14381 | 17 | 17.18 | 71.77 | 21.77 | 2202.69 | 17 | 16.82 | 96.57 | 12.44 | 670.63 | 9 | 4.00 | 98.73 | 1.01 | 145.77 |
| ac97_ctrl | 2199 | 9208 | 21 | 15.10 | 75.05 | 25.05 | 2225.54 | 21 | 15.43 | 94.92 | 10.79 | 800.31 | 8 | 13.00 | 97.73 | 0.01 | 111.38 |
| pci_bridge32 | 3321 | 12494 | 32 | 13.84 | 73.66 | 23.66 | 5124.25 | 32 | 9.41 | 96.76 | 12.63 | 2594.26 | 8 | 9.50 | 98.67 | 0.95 | 586.74 |